# HENRI BÉNARD: THERMAL CONVECTION AND VORTEX SHEDDING

José Eduardo Wesfreid[*]

*Physique et Mécanique des Milieux Hétérogènes, CNRS, ESPCI, PSL Research University,*
*10 Rue Vauquelin, 75005 Paris, France;*
*Paris-Sorbonne Université, 1 Rue Victor Cousin, 75005 Paris, France;*
*and Université Paris Diderot, 5 Rue Thomas Mann, 75013 Paris, France*

---

[*] wesfreid@pmmh.espci.fr

ABSTRACT

We present in this article the work of Henri Bénard (1874-1939), French physicist who began the systematic experimental study of two hydrodynamic systems: the thermal convection of fluids heated from below (the Rayleigh-Bénard convection and the Bénard-Marangoni convection) and the periodical vortex shedding behind a bluff body in a flow (the Bénard-Kármán vortex street). Across his scientific biography, we review the interplay between experiments and theory in these two major subjects of fluid mechanics.

1. INTRODUCTION

The name of Henri Bénard is associated with two of the most active research subjects in fluid dynamics: the thermal convection of fluids heated from below, called the Rayleigh-Bénard convection (Figure 1a) and the vortex shedding behind a cylinder, called the Bénard- Kármán vortex street (Figure 1b). He was the scientist who began the systematic experimental study of these two systems at the beginning of the 20th century[1].
These experiments opened a large theoretical and experimental research field, especially in the last 40 to 50 years. Indeed, they are strongly connected with major advances in the study of turbulence, chaos, instabilities, patterns, nonlinearities or bifurcations theory [2]. In addition, spatial patterns displaying order and disorder structures in convective instabilities, inspirit analogies outside of the original field of study, for example in social sciences, around the concept of dissipative structures [3]. Through his scientific biography, we will learn about the evolution of these main subjects of research and the interplay between pioneering experiments and theory. We will also take a quick look at the development of fluid mechanics laboratories in the interwar years in France.

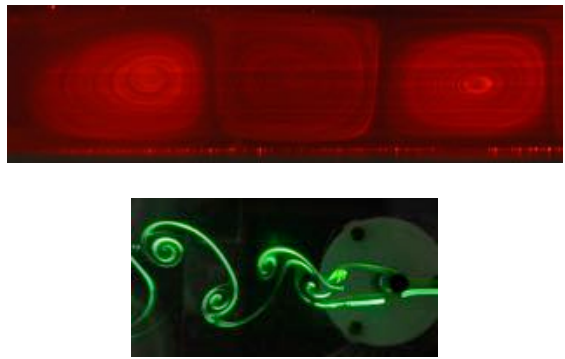

Figure 1: a) Rolls in Rayleigh-Bénard convection; b) Bénard- Kármán vortex street

---

[1] He wrote in his resume [1]: *Deux grands travaux d'hydrodynamique expérimentale ont occupé la plus grande partie de ma vie scientifique: deux problèmes, d'ailleurs très différents, m'ont amené à découvrir les éléments géométriques simples, très bien définis, dans deux phénomènes très répandus, très fréquemment observés dans la nature, mais jusqu'alors très négligés des physiciens. D'heureux hasards m'ont évidemment servi. En parcourant avec moi l'historique détaillé des tourbillons cellulaires, qui constitue le premier travail, et celui des tourbillons alternés, derrière un obstacle mobile, qui constitue le second, on sera frappé, je crois, du fait que, dans les deux cas, j'ai à défricher un terrain presque vierge. Toutefois, dans l'un et l'autre cas, j'ai ignoré un premier explorateur antérieur à moi: James Thomson, pour les tourbillons cellulaires; A. Mallock, pour les tourbillons alternés. Bien que je n'aie pas utilisé les entiers qu'ils avaient tracés, les ayant connus beaucoup plus tard, je suis heureux que cette Notice me donne l'occasion de mettre leur rôle de précurseurs en évidence (ce que j'ai d'ailleurs déjà fait dans mes publications*)

## 2. BIOGRAPHY [4]

Henri Claude Bénard was born on October 25th, 1874 in Normandy, France. After attending elementary school near his birthplace, he moved to Paris to study at the Lycée Louis le Grand (secondary school). In 1894, he passed the very hard entrance exam for the sciences section of the Ecole Normale Supérieure in Paris where he shared the classrooms with only sixteen other students, among whom Paul Langevin and Henri Lebesgue (Figure 2).

In 1897, Bénard became a laboratory assistant ("préparateur"), working in optics with Eleuthère Mascart (Chaire de Physique Expérimentale) and in hydrodynamics with Marcel Brillouin (Chaire de Physique Générale et Mathématique). With Brillouin he had the opportunity to repeat classical experiments from Gotthilf Hagen, Osborne Reynolds and Maurice Couette on transition to turbulence [5-6]. Simultaneously, he was preparing his PhD thesis on the convective motion of liquids. During this work, using optical methods, he discovered the spatial pattern or cellular organization of this kind of fluid motion. On March 15, 1901, he defended his thesis [7].

In 1901, he obtained a short-time fellowship from the Foundation Thiers and married Clémentine Malhèvre. In April 1902, Bénard was appointed assistant professor at the Faculty of Sciences at the University of Lyon, beginning a new experimental activity of research around the observation of free surface deformation behind an elongated vertical body moving in a container filled with liquid. He associated this deformation with the presence of alternating vortices in the fluid.

In April 1910, he took a position as professor of general physics at the Faculty of Sciences of the University of Bordeaux and joined Pierre Duhem's laboratory [8]. After 1914, during the World War I, he served as an officer in the Superior Commission of Inventions of the War Office and later became the Head of the Physics Section under Jules-Louis Breton's direction [9].

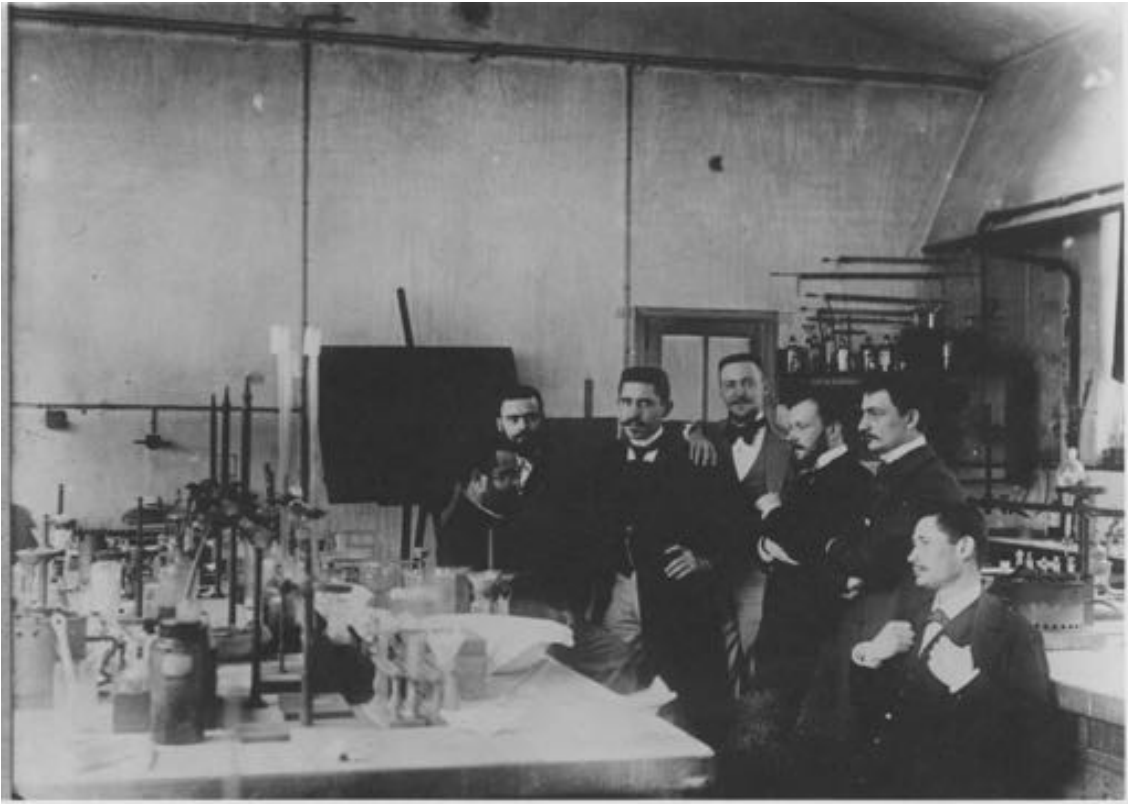

Figure 2: Picture of Bénard and his classmates at the Ecole Normale Supérieure in 1896. From left to right: Jean-Baptiste Perrin (laboratory assistant and future Nobel Prize in 1926), Louis Dubreuil, Paul Langevin, Lucien Patte, **Henri Bénard**, François Meynier and Pierre Maussolier

In November 1922, Bénard moved from Bordeaux to Paris, where he was appointed PCN professor of physics for the first year of medical studies[2] at the Faculty of Science of Sorbonne University and, in 1926, he was named non-tutorial chair professor. In 1928, he was elected President of the French Physical Society for one year, after Louis Lumière and before Jean Perrin [10].

In 1929, he became full Professor and head of the Laboratory of experimental fluid mechanics (Laboratoire de Mécanique Expérimentale des Fluides) at the new Institute of Fluid Mechanics of Paris (Institut de Mécanique des Fluides de l'Université de Paris–Fondation du Ministère de l'Air), headed by Henri Villat. The Institute was created following an initiative from the Air Ministry, which, under Albert Caquot's coordination [11], also created about the same time other similar institutes in various cities across France [12].

---

[2] *PCN* (certificat propédeutique aux études médicales)

Henri Claude Bénard died on March 29, 1939 at Neuilly-sur-Seine at age of 64. After his death, Adrien Foch was named as his successor as Professor of Fluids Mechanics at the Institut de Mécanique des Fluides. Meanwhile Yves Rocard moved from Clermont-Ferrand to Paris to become assistant professor in this chair.

3. THERMAL CONVECTION WORKS

When Bénard was a laboratory assistant at the Collège de France, trying to prepare a coherer with solid dielectrics, he observed, in a melten paraffin bath, that the motion of graphite particles formed a pattern of "semi regular polygonal" cells. Interested in this organisation, and using optical methods, he then studied the movement of the particles in a thin layer of fluid heated from below with an upper free surface inside a copper vessel with very well controlled temperature, paying special attention to surface deformations occurring because of the convection.
On March 15, 1901, he defended his PhD thesis on the convective motion of fluids and published at the same time major papers including a full description of convective motion, especially the hexagonal pattern of the convective cells, today called "Bénard cells" or "Bénard hexagons" (Figure 3 and 4b). From the detailed description of the closed trajectories of particles inside such cells, with hot liquid ascending around its axis and descending along the vertical sides at the periphery (Figure 3), he referred to these in terms of cellular vortices ("tourbillons cellulaires").

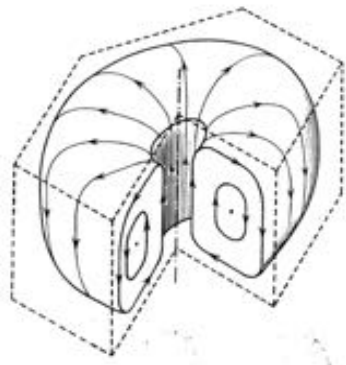

Figure 3: Schema of the hexagonal cellular vortex [7]

As a good experimentalist and a fine physicist, he designed his experiment to insure extreme homogeneity in the hot planar bottom of the bath containing the fluid and its temperature. Indeed, he wanted to eliminate "the smallest fluctuation or local excess of temperature… sufficient to create a centre of ascension"[1].

All these experiments of Bénard's were performed with a free surface, using different fluids such as spermaceti and melted paraffin wax. He used various methods of observation in order to obtain very precise measures of the size of the cells (of the order of magnitude of two times the thickness of the fluid layer), from direct photographs to follow the trajectories of graphite powder in suspension in the cells to shadowgraphic optical methods of refraction and transmission in thin layers. These results, summarized quantitatively in Figure 5, will play a role in the future discussions about the physical mechanism behind fluid organisation.

Bénard also observed other patterns of motion such as "tourbillons en bandes" or convective rolls of constant spacing.

Before Bénard, cells with convective motion were described only in Ernst Heinrich Weber's work [13] on the evaporation of a thin layer of alcohol and also by James Thomson [14], who studied the slow cooling of a warm film of soapy water showing a tessellated structure of convective circulation (Figure 4a). Bénard was not aware of these previous publications (see footnote 1). His rigorous and systematic procedures of observation and measurement, obtained with a very homogeneous and controlled heating, allowed him to discover a highly regular and extensive pattern of fluid motion, never obtained before.

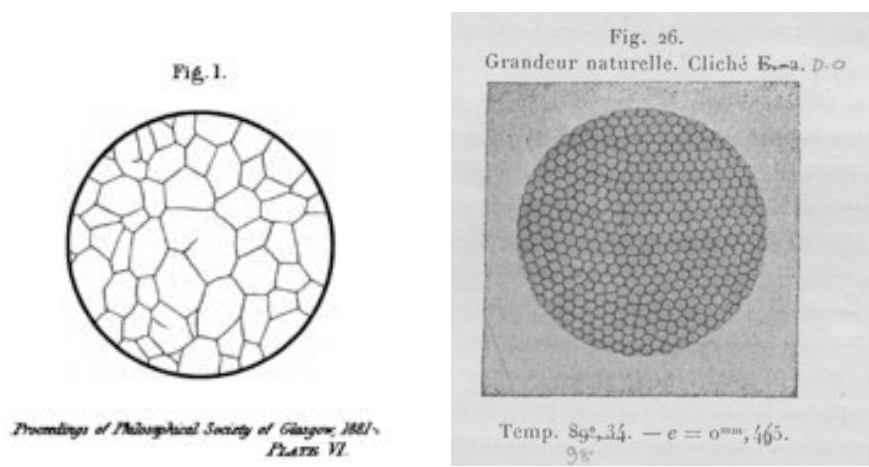

Figure 4: a) prismatic cells observed by Thomson[14]; b) cellular pattern observed by Bénard[7]

3.1. RAYLEIGH-BÉNARD CONVECTION

In 1916, Lord Rayleigh published a paper, one of the last one written before he died, entitled "On convection currents in a horizontal layer of fluid, when the higher temperature is on the under side" [15], where he wrote: *The present is an attempt to examine how far the interesting results obtained by Bénard in his careful and skilful experiments can be explained theoretically*. Rayleigh obtained, by linear stability analysis of the fluid heated from below, the theoretical conditions for the formation of convective cells in the form of rolls, with a rather artificial situation of no-slip (or stress-free) horizontal boundaries. He considered that the buoyancy, induced by thermal expansion, produced the convective motion, when the non-dimensional difference of temperature $Ra = \beta g (T_b - T_u) h^3 / \nu \kappa$; (defined[3] in [16-17] and later called Rayleigh number in [18]) across the fluid layer exceeds a critical value $Ra_c = 27/4 \pi^4 = 657.51$. In addition, he predicted that the wavelength $\lambda$ of the cellular periodicity is given by $\lambda = 2h$

A layer of fluid, in presence of buoyancy induced by difference of temperature, is stable if cooled from below, always unstable if any horizontal gradient of temperature exists, and unstable when heated from below, if the Rayleigh number exceeds its critical value $Ra_c$. This last situation is what is named the Rayleigh-Bénard convection.

Bénard read Rayleigh's paper after the war, in 1920[4], and then compared the theoretical prediction of the distance *l* between cells, with the results obtained in different experiments during his thesis, which were performed in a very thin layer of thickness *h*, of about 1 mm, almost displaying hexagonal cells. These results show a length *l*, at the onset of deformation, equal to *0,280h*, as we can see in the Figure 5.

---

[3] $Ra = \beta g (T_b - T_u) h^3 / \nu \kappa$; where $T_u$ is the temperature of the top plate, $T_b$ is the temperature of the bottom plate, $g$ is the acceleration due to gravity, $h$ is the thickness of the fluid layer, $\nu$, $\kappa$ and $\beta$ respectively its kinematic viscosity, thermal diffusivity and thermal expansion coefficient.

[4] He later relates in his resume [1]: *Ce Mémoire, paru en pleine guerre, m'a d'abord échappé; les physiciens des nations béligérantes ne lisaient pas tout ce qui les intéressait. Je ne l'ai connu qu'après la mort de lord Rayleigh, que je regrette de n'avoir pu remercier de sa trop élogieuse appréciation de mon travail experimental.*

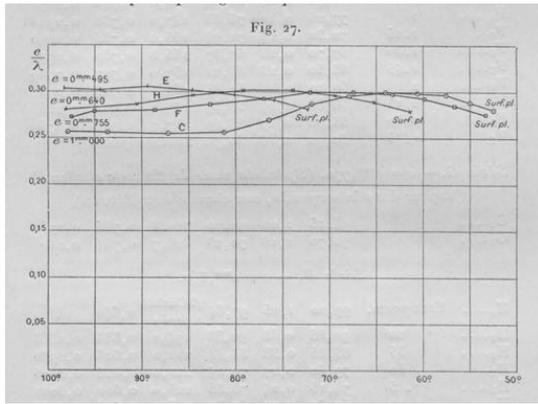 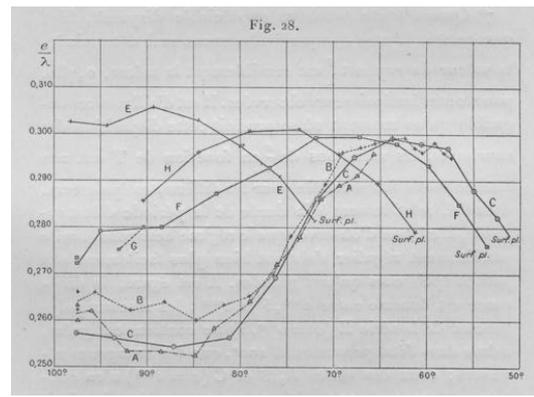

Figure 5: a) Variation of the ratio between size *l* of the hexagonal cells and thickness *e* (*h* in our text) of the fluid layer, as a function of the temperature during cooling, for experiments at different thickness; b) zoom of the results, near the onset of convection

In 1927 and 1928 [19-21] Bénard claimed, for the first time, the satisfactory agreement of his measured values of the hexagonal cells' size, as predicted by Rayleigh, when estimated in comparison with the circle of the same area, giving *l* ≈ *0.286h*. In addition, his few results showing square cells are also favourably compared with another one of Rayleigh's prediction for its size, even if Bénard noted the unrealistic free slip boundary conditions for two isothermal surfaces used in the theoretical analysis.

New theoretical studies of linear stability, following Rayleigh's, were obtained by Harold Jeffreys [16-17] and Major A. R. Low [22]. In the case of a rigid boundary at the bottom wall containing the fluid and a free boundary at the top, as in Bénard's experiment, Jeffreys obtained a higher critical value of $Ra_c$ = 1051 in 1926 and 1928. It should be noted that, in all this work, explicit reference is made to Bénard's experiments, very well known within the scientific community and performed nearly 30 years previously. But on the other hand, no attempt was ever made in the very same work, to compare the predicted critical values of the difference of temperature which were observed in the experiments, at the transition point between the hexagonal cells and the stable motionless plane surface in the fluid layer (noted *Surf. pl* in the Figure 5).

Bénard, in his 1928 paper [21], focused his attention on the spatial scales *l* of the convective cells, leaving out the question of the onset of transition and in particular the value of the critical Rayleigh number[5].

In 1934, during the course of two conferences, at the 4th. Int. Congress of Appl. Mech, in Cambridge and in the Journées de Mécanique des Fluides in Lille, Bénard presented his analysis of old data, estimating the critical temperature of extinction of convection. Compared with Jeffreys' calculations, done within the same kind of boundary conditions as the experiments (rigid-free b.c), he obtained for his experimental results to be 300 to 1000 times smaller than predicted at the onset! Bénard spoke therefore of "the Rayleigh deficiency".[23]

At that time, other studies of thermal convection with air and other gases, almost always motivated by the analogy with clouds patterns and in conditions where the fluid layer is between two horizontal rigid plates of good thermal conductivity, were performed in England [24] as well as in Bénard's new laboratory at the Institute de Mécanique des Fluides at the Sorbonne mainly with Victor Volkowsky [25] and Dusan Avsec[6][26]. During those studies, deviations were noticed with the values at the onset predicted by the Rayleigh-Jeffreys' theory, but explained at that time by the existence of a non-linear temperature gradient as well as by variations of viscosity with temperature (called non-Boussinesq effects). This subject conformed an important part of Avsec's PhD thesis. But the magnitude of these deviations was not as important as Bénard claimed in his experiments, performed with the upper free surface.

Finally, in 1935, R. J. Schmidt and S. W. Milverton [27], in a very well controlled experiment with water heated from below between rigid plates, observed the modification of the heat transfer at the point of transition between convection and conduction regimes, at values of the Rayleigh numbers $Ra_c$, confirming Jeffreys' calculations for these boundary conditions.

---

[5] Bénard wrote: *Bien qu'il (Rayleigh) déclare expressèment avoir voulu expliquer mes anciens résultats expérimentaux, il n'a pas poussé jusqu'au contrôlé des valeurs numériques. Or,* **au moins en ce qui concerne les dimensions transversales des cellules et des bandes, l'accord est excellent**.
*(Par rapport à la condition nécessaire pour qu'il y ait équilibre stable préconvectif d'une lame liquide chauffée par en dessos)........**je laisserai de côté pour le moment** le contrôle expérimental de ces limites du régime d'équilibre stable préconvectif.* [21]

[6] Dusan Avsec (1905-1989), a Slovenian student who arrived in Paris in 1934 and spent 5 years with Bénard, working on applications of convection to atmospheric fluid mechanics. Following previous activity of Pierre Idrac (1885-1935) on thermal convection under shear, producing longitudinal rolls similar to the clouds streets, he obtained support of the Atmospheric turbulence committee of the French meteorological service. In 1940, he returned to Ljubljana, where later, he becomes head of the Engineering Faculty.

3.2. BÉNARD-MARANGONI CONVECTION

For half of a century, Lord Rayleigh's explanation, considering buoyancy as the physical mechanism at the origin of the cellular convection described by Bénard, was widely accepted. A turning point occured when in a very short paper published in *Nature* in 1956 [28], Myron J. Block, a scientist from a private laboratory, Baird Associates Inc., described the repetition of the original Bénard's experiments, in a very thin layer of a hydrocarbon liquid, observing the hexagonal cellular structure. Spreading a thin layer of insoluble silicon on the film, he observed that the deformation of the free surface and the convective circulation stopped immediately, and said that "the variation of vertical density could not be affected so rapidly" and, using results from A. V. Hershey, he suggested that "the Bénard cell motion and deformation are produced by variations of the surface tension, which are in turn due to variations in temperature". So he regarded the Marangoni effect as the motor of the fluid motion and "it would therefore appear that the instability causing Bénard cells is quite different from the Rayleigh instability".
Two years later, Anthony Pearson [29], as postdoctoral researcher in the industry, repeated the full stability analysis of Rayleigh, considering a layer of fluid heated from below and heat loss from the upper free. Motivated by observations on drying paint[7], he considered no buoyancy term but introduced another force term, the Marangoni force, related to the variation of the surface tension with temperature (or concentration), but without reference to Block's[8] previous work. He used as non-dimensional control parameter, the Marangoni number *Ma*, which is linearly proportional to the thickness of the layer[10] and obtained a critical value of instability $Ma_c$ = 80 and a critical wavelength $\lambda_c = \pi h$. Further work from Donald Arthur Nield [30] included buoyancy term like Rayleigh's and gradients of surface tension term like Pearson's, combining these rival theories. It has been found that the two mechanisms causing instability reinforced one another and were tightly coupled[11], with critical behaviour given by the condition $Ra/Ra_c + Ma/Ma_c = 1$. As *Ma* increases with *h* and *Ra* with $h^3$, it might be expected that the influence of the surface tension becomes more important for very thin layers of liquid, as used by Bénard (and Block). The critical Marangoni number $Ma_c$, was almost certainly exceeded and convective motion appeared in these experiments. In his experiments, the measured Rayleigh number R was 300 to 1000 times smaller than the $Ra_c$ number predicted by Rayleigh and Jeffreys. The strong disagreement between the experimental results and the existing theories[12] explain Bénard's sceptical attitude in 1934 [23] around the validity of the existence of a critical temperature difference.

---

[7] In his paper [29], Pearson described this motivation as follows: *Drying paint films often display steady cellular circulatory flow of the same type as that observed in the case of fluid layers heated from below. In the latter case (that of the so-called Bénard cells) the motion can usually be ascribed to the instability of the density gradient that would be present if the fluid were stationary. This cannot be the mechanism causing the flow in the former case, since the circulation is observed whether the free surface is made the underside or the topside of the paint layer, that is, even if the gravity vector is effectively reversed.*

[8] In a private communication (Cambridge, 2017), Pearson remembers that: *Dr Cousins was a physicist working for ICI at their Paints Division at Slough, whom I visited when I had just re-joined the Company at their (special fundamental) Akers Research Department, The Frythe, Welwyn, Herts. I was intent on showing that Mathematical Modelling* (the former subject of Pearson in Cambridge) *could be of importance to the company, and so visited all their Divisions to learn about the problems they needed to*

But the theoretical predictions on the cells size, when the Marangoni effect is included, are approximately the same as those obtained with the Rayleigh-Jeffreys theories when only buoyancy is included. This explains the positive comparisons made by Bénard [19-21] in 1927-1928, between his results shown in Figure 5 and Rayleigh-Jeffrey's theories.

---

*solve. He had already realised that Bénard's theory could not explain a phenomenon that he had noticed in their green paint: when painted on a test slide, the two pigments separated into a cellular pattern whether the paint was applied to either the upper or the lower surface of the slide held horizontal. He had not thought that surface tension might be the cause.*

*… I owe Block an apology for not citing his paper, even though at the time I was unaware of his paper and so was Dr Cousins (Block wrote me a rather aggressive letter pointing it out to me!) and searching the literature was not as easy as it is now.*

*…Industry is still the best source of real problems in continuum mechanics.*

[10] The Marangoni number, used originally in the chemical engineering literature, is defined as $Ma = (d\sigma/dT)(T_b - T_u) h/\nu\kappa$ where $T_u$ is the temperature of the upper surface, $T_b$ is the temperature of the bottom plate, *h* is the thickness of the film, $\nu$ is the kinematic viscosity, $\kappa$ is the thermal diffusivity and $(d\sigma/dT)$ is the coefficient of variation of the surface tension $\sigma$ with temperature.

[11] Buoyancy must, of course, be the only mechanism concerned when there is no free surface.

[12] Pierre Vernotte, working with Bénard, suggested another theoretical explanation to this disagreement, based on different relations between buoyancy and inertial terms, but remainder too general[33].

Later, further developments in nonlinear theories [31] show that the initial preferred patterns are hexagons, in the case of surface tension induced convection (Bénard-Marangoni convection in modern language), and rolls when buoyancy is responsible for the instability[13] (Rayleigh-Bénard convection, as called today). This transition was precisely what Bénard and Camille Dauzère observed in an experiment when a thin layer, heated from below, is covered with solid wax and begins to melt. (Figure 6). One of Bénard's most refined experimental results was the measurement, with very precise optical methods, of a very small depression in the centre of the cells by about 1 μm, where up flow exists. The direction and amplitude of this free surface deformation was explained in the 60's by Laurence Edward (Skip) Scriven and Chuck Sternling [32]. Bénard himself called attention to the role of surface tension, when he analysed this observation[14].

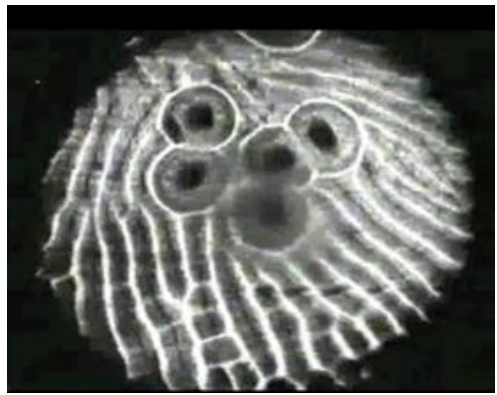

Figure 6: Experiment of thermal convection in a layer of fluid covered by paraffin on the top and partially free in the centre, where few hexagonal cells develop, while, in the covered film, are rolls visible ("tourbillons en bandes"). Picture taken from the film by Dauzère and Bénard (1913)

His pioneering experiments on thermal convection and patterns motivated further deep theoretical and modern experimental work. A full overview of this evolution can be found in E. Lothar Koschmieder's book [34] where he also details the study of the wavelength selection obtained experimentally by Bénard and explained in the 1980's.

---

[13] Except if $\nu$, $\kappa$ and $d\rho/dT$ change across the layer (Non-Boussinesq convection)

[14] He noted that: *La tension superficielle, à elle seule, provoque déjà une dépression au centre des cellules et un excès de pression sur les lignes de faîte qui séparent les cuvettes les unes des autres.*

## 4. VORTEX SHEDDING EXPERIMENTS

The young Bénard was given his first position at the University of Lyon, after his remarkable PhD. Backed by his experience in optical methods in hydrodynamics[15], especially in measurements of micrometric deformations in free surfaces, he tried to apply these techniques to new studies in experimental hydrodynamics.

With these motivations, in 1904, he built, in a cellar in the Faculty of Sciences, a water tank where a vertical object, a dessert knife partially immersed, is displaced producing a deformation in the free surface. He reported[16] that he "was surprised to discover two splendid lines of alternate vortices, a phenomenon that many people can see everyday without noting it" [1]. These vortices reminded him of the eddies or sinuosity discussed in Brioullin's lectures [18] at the Collège de France, on the transition between the laminar state (called "Poiseuille regime") and the turbulent one (called "hydraulic regime").

In 1906, the experiment was operational, with control of the temperature of the fluid as well as of the uniformity of the motion of the obstacle.

He studied the small deformation of the free surface produced by the vortex, with the optical Foucault knife-edge method. The double line or street of alternate vortices develops small mirrors reflecting the light. He perceived images in the form of a crescent at low velocity of the obstacle and in the form of a scythe at higher velocity, with nearly grazing light (Figure 8b). He observed 20 or 30 vortices visible to the naked eye. Pictures were taken to characterise the geometry and displacement.

The first experiments were not convincing because of the difficulties of the synchronisation between the motion of the obstacle and the automatic system taking pictures. So a few months after the first experiments, he turned to the new cinematography technique, using the apparatus distributed by the Lumière brothers in Lyon: a Lumière-Carpentier camera with a motor (Figure 7). Bénard performed systematic experiments using nearly rectangular elongated obstacles, with a small rounded or pointed form in the upstream face. The aspect ratio $L/l$ of these obstacles varied between 2,5 and 40, when $L$ is the length of the body in the direction of the flow and $l$ its transversal width.

---

[15] Bénard described these methods in one of his first papers on cellular convection [35] and gave a full panorama of the importance of visualisation in fluid mechanics in his inaugural address of the Institute de Mécanique des Fluides de Paris, on 13th November 1929 [36]

[16] *il me semblait que tous les problèmes même classiques, même traités complètement par les mathématiciens, aussi bien ceux où la viscosité et la conductibilité thermique ne jouent qu'un rôle négligeable, que ceux où ces deux propriétés sont d'importance prépondérante, devaient être repris par les physiciens, en y appliquant toutes les ressources de l'optique, de la photographie et du cinématographe. C'est avec ce beau dessein que je commençai à remuer une cuiller à café à la surface d'un cristallisoir, plein d'eau chargée d'encre, pour répéter les expériences de Helmholtz sur les « tourbillons du café au lait ». Mais je variais beaucoup la forme de l'obstacle* [1].

[18] Marcel Brillouin, after two years as Mascart's substitute, was named Professor at the College de France in November 1900.

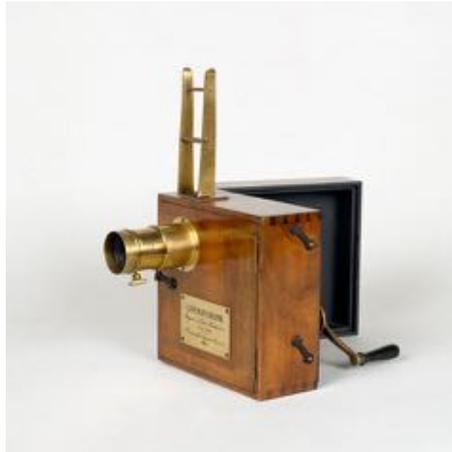

Figure 7: Cinematographic camera Lumière- Carpentier as that used by Bénard, adding a motor

Bénard announced his results in November 1908 [37-38] in the Comptes rendus de l'Académie des Sciences, with two short papers essentially centred on the spatial properties of the alternate vortex distribution. He observed that the wavelength $\lambda$ (distance between successive vortices in one line) growed with the transversal dimension $l$ and did not depend on the longitudinal one $L$. It also increased with the viscosity $\nu$ of the fluid but not, in first approximation, with the velocity $U$ of the obstacle[19].

These publications had a strong impact and were cited by Théodore von Kármán in his 1911 and 1912 [41-42] papers on the stability of the vortex street. They were also referred to in two papers on Aeolian sounds, one from Friedrich Krüger and Adolf Lauth in 1914 [43] and another one from Lord Rayleigh in 1915 [44]. They analysed Vincenz Strouhal's experiments [45] on the production of sound by wires in the wind (Aeolian tones) and explained this effect by the shedding of alternate vortices observed by Bénard.[20].

---

[19] He recognized later, that he has written a bit hastily and made an error in drawing an inverted rotation direction of one line of vortex [39]. In addition, in 1926 [40] he showed how $\lambda$ change with the velocity.

In 1911, Marcel Brillouin also recognised Bénard's results as a confirmation of the waviness or vortex rolling that he described in his book [5] as being present in turbulent situations [21]

---

[20] In 1915, Lord Rayleigh gave in this paper[44] a synthetic view on the state of the art of the problem:
*As regards dynamical explanation it was evident all along that the origin of vibration was connected with the instability of the vortex sheets which tend to form on the two sides of the obstacle, and that, at any rate when a wire is maintained in transverse vibration, the phenomenon must be unsymmetrical.*
*The alternate formation in water of detached vortices on the two sides is clearly described by H. Bénard: "Pour une vitesse suffisante, au-dessous de laquelle il n'y a pas de tourbillons (cette vitesse limite croît avec la viscosité et décroit quand l' épaisseur transversale des obstacles augmente), les tourbillons produits périodiquement se détachent alternativement a droite et a gauche du remous d'arrière qui suit le solide; ils gagnent presque immédiatement leur emplacement définitif, de sorte qu'à l' arrière de l'obstacle se forme une double rangée alternée d'entonnoirs stationnaires, ceux de droite dextrogyres, ceux de gauche lévogyres, sépares par des intervalles égaux."*
*The symmetrical and unsymmetrical processions of vortices were also figured by Mallock* (see Figure 8 ) *from direct observation. In a remarkable theoretical investigation, Karman has examined the question of the stability of such processions. The fluid is supposed to be incompressible, to be devoid of viscosity, and to move in two dimensions. The vortices are concentrated in points and are disposed at equal intervals λ along two parallel lines distant h. Numerically the vortices are all equal, but those on different lines have opposite signs. Apart from stability, steady motion is possible in two arrangements (a) and (b), fig. 1, of which (a) is symmetrical. Karman shows that (a) is always unstable, whatever may be the ratio of h to l; and further that (b) is usually unstable also. The single exception occurs when $\cosh(\pi h/\lambda) = \sqrt{2}$, or $h/\lambda = 0.283$. With this ratio of $h/\lambda$, (b) is stable for every kind of displacement except one, for which there is neutrality. The only procession which can possess a practical permanence is thus defined."*

[21] Brillouin wrote : *La formation de ces tourbillons par une impulsion à l'arrière d'un obstacle, ou même sans obstacle, à la sortie d'un orifice est connue depuis longtemps et très facile à observer. Mais dans un courant permanent, l'observation est moins facile. Toutefois, une installation imitée de celle de Marey, où l'on observe à l'aide de filets de fumée, le mouvement de l'air au voisinage d'un obstacle dans un large tube à section rectangulaire (40 cm de côté), m'avait conduit, en 1903, à la conviction que le phénomène régulier dans un fluide naturel est, non pas la formation d'une surface de discontinuité permanente, mais la production périodique de nappes de discontinuité qui s'enroulent par leur bord en spirale de plus en plus serrée, se transforment en tourbillons réguliers, sous l'influence de la viscosité, et qui, toujours sous cette même influence, s'étalent et finissent par s'éteindre, au sein du liquide qu'ils accompagnent dans son mouvement général. Cette conviction née d'observations multipliées, mais seulement qualitatives, a reçu des expériences précises de M. Bénard une précieuse confirmation.*

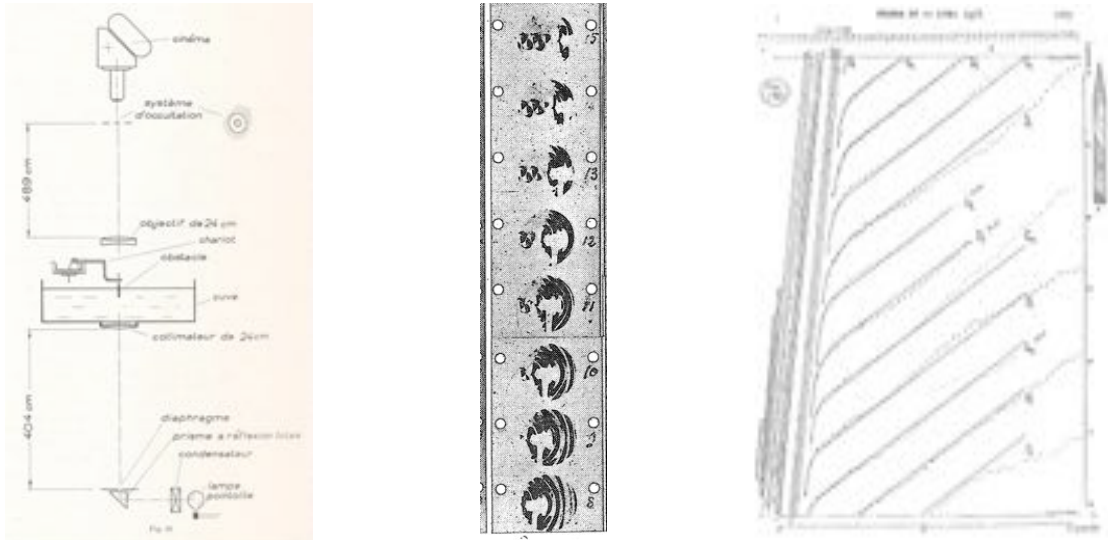

Figure 8: a) schema of experimental apparatus, with the water tank and optical system for the cinematograph (here in the version of the facility that was installed in Paris in 1930); b) sequence of the original films[22] showing the images of the vortex shedding during the motion of the body to the right. In image 8, ripples of the fluid surface around the body are observed. In image 10, on the left, a first vortex appears within the vision field and in image 15, when the body leaves the vision field, the two lines of alternate vortices clearly appear; c) spatio-temporal plot of each vortex displacement motion. On the left hand side, is plotted the displacement of the bluff body. The difference of slopes gives the phase velocity of the shedding.

In 1910, when Bénard moved to Bordeaux, the experiment, too big and too complex, was dismantled. But still, until 1925 he continued to analyse under microscope close to 20000 frames of 133 films obtained at Lyon, in order to observe the micrometric displacement of the images of the vortex shedding. In 1913 [46], from a more accurate observation of the films, he published very precise spatio-temporal diagrams of the vortex shedding showing that the vortex moves more slowly that the body ($c<U$) , as reproduced in Figure 8c. The data processing from his films was discontinued during the war[23].

---

[22] Note the circular perforation on each side of the image, following Lumière's patent. Later the Lumières adopted the American system of Edison, with four rectangular perforations.

[23] During two years Bénard served as sergeant in Bordeaux, working as a scientist on the problem of thermal insulation of wagons transporting frozen meat for the Army. This work led to a new method of measurement of the thermal conductivity, using a periodic signal to check the insulation, inspired by the original work of Fourier on the seasonal propagation of the heat within the Earth's soil.
This work was performed with the collaboration of Pierre-Michel Duffieux, who later introduced the Fourier techniques in optics. In an autobiographic text [52], Duffieux recognised that Bénard, his "military chief ", influenced him to learn about Fourier transform. In 1916, Bénard moved to Toulon,

Owing to Brillouin's insistence and the impact of his previous notes to the Comptes rendus, he came back to the study of the original films, choosing 77 of those which presented a more regular vortex shedding. In 1926, he published four new notes in which he exposed his original experimental results in the context of the theoretical results from Kármán and Hans Ludwing Rubach [42] and Lord Rayleigh [44] as well as new experiments from Ernest Frederick Relf [47-48], Edward Gick Richardson [49] and Charles Camichel's group, in Toulouse, group including Pierre Dupin and Maximilien Teissié-Solier[50-51].

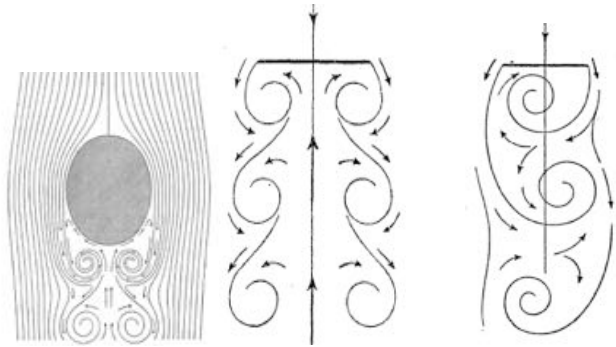

Figure 9: Drawings of Arnulph Mallock[53-54] of symmetric and alternate vortices. The one on the right erroneously describes a single helicoidally vortex where the alternate vortex is the manifestation of the section plane.

As we mentioned before (footnote 18), Kármán studied the necessary conditions to maintain in a stable position an unconfined and infinite alternate point vortex street of inviscide flow. His theory[41], predicts as a condition, a value of $h/\lambda$ = 0,283, where $h$ is the distance between the two lines of periodic vortex with wavelength $\lambda$. He also predicted the velocity of the vortices, of intensity $J$, as $c = (J/2\lambda)\, th(\pi h/\lambda)$ and the formula for the resistance that a body, moving itself with a uniform velocity in a fluid, must feel. This resistance is calculated according to two experimental parameters: the ratio $a= c/U$ given by the velocity of the vortex $c$ in respect to the fluid velocity $U$ and the non-dimensional wavelength $b= \lambda/l$., where $l$ is the transversal size of the bluff body. These values are related to the frequency of value $f=(U-c)/\lambda$, or in terms of the Strouhal number $S$ ($S = f\,l/U$ is the non-dimensional frequency $f$), $S=(1-a)/b$. It is clear that the sole knowlodge of the ratio $h/\lambda$ is not

---

working with the Navy and the Superior Commission of Inventions of the War Office, on the development of optical instruments of observation.

sufficient, in the frame of this theory, to predict the frequency[24].

Rubach, his co-author, performed an experiment in Göttingen[42] with a body at rest, using a cylinder or a plate, in a stream of water with two different velocities (corresponding to Reynolds number 1705 and 2370 in the cylinder case) and measured these two parameters directly from pictures. He observed that they did not change with the velocity.

In addition, he noted that the measure of the ratio $h/\lambda$ agreed with Kármán's theoretical prediction, but only when measured away from the vicinity of the bluff body.

Bénard's experiments were more precise, measuring all these ratios, covering a wide range of *Re* numbers, in four different liquids and using prismatic bluff bodies with four different sizes of transversal dimension *l* of 0.1, 0.2, 0.4 and 0.8 cm.

We found in his extended resume of 1929 [55], the full experimental data from 71 films, covering the range of $S \in [0,096-0,250]$ and $Re \in [88, 1142]$, which we plotted, on the Figure 11 at the next section.

The obtained non-dimensional frequencies, which, in these experiments, are well defined as a global parameter in the full field, which varies in the range referred to before. They differ from the constant value for *S* obtained from the visualisation of Rubach in Göttingen [42].

Bénard also noted the variation downstream of the ratio $h/\lambda$ between 0.23 and 0.59 in rough agreement with the theoretical expectation, but with strong fluctuations in the estimation of *h*. This ratio is not easy to measure precisely, as indeed, the vortex street expands near the obstacle, giving a higher value for this ratio.

On the basis of these experiments, he supported a strong statement that "the Kármán's law is far from being applicable to vortices in **real**[25] liquids" [40] as it is established for inviscid fluids[26]. He also noted that the streamwise progressive decay of the vortices had not been considered[27].

---

[24] In fact, the Strouhal number that can be deduced from this theory, as *S= (1/b)(1 - J /(bνRe√8))*, if we use non-dimensionalised velocity and the Kármán value for *h/λ*.

[25] We underscore the word real.

[26] Kármán, aware of this problem, refers to the theory of Ludwig Prandtl on separation. In his paper of 1912, he noted « *Much more difficult appears the extension of the theory in another direction, which really would first lead to a complete understanding of the theory of fluid resistance, namely, the evaluation by pure calculation of the ratios a=λ/d and b=c/U, which we have found from flow*

After Kármán's publication, other authors came back to his theory and to the prediction of the parameter $h/\lambda$ and the vortex velocity $c$ such as George Jaffé [57] in 1920 or Werner Heisenberg [58] in 1922, who discussed the case of the vortex street behind a plate. Others, like Hermann Glauert, Susumu Tomotika, Louis Rosenhead [59-60-61] and in France, Villat [62], considering "more realistic" conditions, treated the transversal confinement using image vortices. They predicted a range of values for $h/\lambda$ as a function of the aspect ratio of the channel, with 0,283 as an asymptotic value for unlimited fluids. The calculations became more complex but were still remote from the experiments. This generated strong comments, as in a recension of a Villat's book [63] devoted nearly entirely to this problem, in *Nature* where it is expressed[64] "that the author cares much for mathematical analysis and little, if at all, for real fluids"[28]. Henri Bouasse, a prolific French physicist, known for fierce comments, said "the problems of stability, such as Kármán on the alternate street, are elegant exercises of calculus devoid of physical sense" [65].

## 4.1 SIMILARITY LAWS

In the series of papers of 1926, Bénard discussed the fluid similarity enounced by Rayleigh[44] in reference to the Aeolian sounds and the experiments of Strouhal. Strouhal in 1878[45], investigating the cause of the musical note heard when a wire of circular section is moved through the air at considerable speed, found that the frequency of the note was independent from the tension of the wire and from its elastic constants, suggesting that the Aeolian sounds came from hydrodynamics and not from elasticity (even if the role of the vortex shedding was, at that time,

---

*observations, and which determine the fluid resistance. ... An apparent contradiction is brought out by the fact that we have used only the theorems established for perfect fluids,.. ...Whether or not this would meet with great difficulties can not at the present time be stated."*

[27] The streamwise variation of the vortices was measured one century later [56].

[28] An anonymous editor wrote: *Prof. Villat's treatment of vortex motion has both the virtues and defects which are usually found in French treatises on mathematical physics. The mathematical treatment is clear and logical, and presented in an attractive style. On the other hand, although the lectures on which the book is based were delivered at an Institute founded by the Ministry of Air, we have scarcely any reference to experimental data. There is one oasis in the desert of mathematical symbols (p. 80), where we read that a cylinder moving in liquid is really found, in certain circumstances, to set up two series of vortices closely conforming to those calculated by Bénard. With this exception, the book suggests that the author cares much for mathematical analysis and little, if at all, for real fluids. However, if we accept his point of view, there can be no question as to the quality of the work* [64]

unknown). He found that the frequency *f* changed with the wind velocity *U* and the diameter *l* of the wires producing a sound, showing that the parameter *f l/U* (later called the Strouhal number *S,* by Bénard) is constant and equal to 0,185.

Rayleigh applied the concept of similarity, in order to estimate the effect of viscosity on the law for the frequency. In 1894, in the second edition of his Theory of Sound, [66], he proposed that the frequency in the Aeolian harp had to be a function of the inverse of the Reynolds number and suggested a general law, from a series development around this small parameter[67]. Later, in 1915, he gave[45] the law *S = 0,195 (1 - 20,1/Re)* from the analysis of Strouhal's results, and of his own experiments in water, and explained these tones with the vortex shedding of Bénard (see footnote 18), as proposed by Krüger and Lauth[29], who observed the variations of the Strouhal number with the velocity.

After Bénard's detailed experiments, other quantitative experiments were performed to test the Rayleigh similarity law explaining the Aeolian tones, which produced new data about the variation of the frequency of the vortex shedding with the velocity.

In 1914 and 1919, Krüger[43-68] observed the variations of the Strouhal number with the velocity. In 1921, Relf was the first to plot the data as a curve S vs. Re, for his experiments in water and in air [47] (Figure 10a). In 1924, Richardson performed experiments in water and air using different methods [49]. Camichel, Dupin and Teissié-Solier published in 1927 [50] a more complete curve *S vs. Re*, including subcritical oscillations for circular cylindrical bluff bodies and finally, in 1928, Bénard [69] compared his own data, obtained, as we explained before, with prismatic elongated bodies in four different liquids, and observed the same tendency as Camichel, but on different curves (Figure 10b). He concluded: "A priori, in absence of a rigorous mathematical theory, it does not seem at all certain to me, that the vortex produced by the bodies as knives, have to follow the same law as those of circular cylindrers of the same thickness *l*". There is a lack of geometrical similitude.

---

[29] Indeed Mallock in his papers of 1907 and 1910 [53-54], after showing drawings of vortex shedding behind a plate (Figure 9), in symmetric and alternate configuration, noted that they *cause (are due to ) the vibrations of the Aeolian harp string* . Kármán in 1912 [42] recognize that Carl Runge, in Göttingen, had already drawn his attention on *the tone that is emitted by a stick rapidly displaced in air is fixed by the periodicity* of the vortex shedding

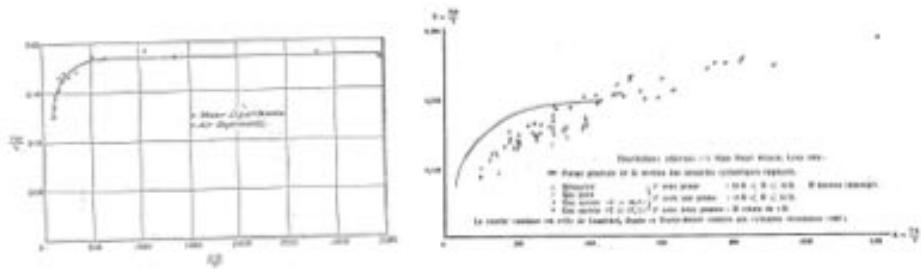

Figure 10 : Similarity curves from the publications of a) Relf[47];
b) Bénard[55], included the Camichel's data.

## 4.2 BENARD - KÁRMÁN INSTABILITY

Questions arose at the interpretation of these experiments, especially Strouhal's and Bénard's and about the theories from Kármán and Rayleigh [70]. These questions were resolved some 60-70 years later, with the theories of instabilities and nonlinear oscillators.

For the application of these concepts, a crucial point was the experimental observation of the existence of a threshold or limit to the laminar state. Bénard, in his 1908 papers [37], noted the existence of a "critical velocity to be exceeded in order to produce the vortices, which grows with the viscosity and decreases when the transversal dimension of the body grows" which definitely, that is a critical Reynolds number $Re_c$. He also referred to the transition from the laminar state to the "turbulent"[30] one revealed by the vortex shedding.

In 1926, he published the values of $Re_c$[71] observed in many of his experiments, displaying some dispersion of these values according of the viscosity of fluids used in each experiment. But he explained that these experiments were not performed in order to obtain $Re_c$, and some times he confounded this minimum $Re$ number with the value of $Re$ when the frequency goes to zero.

In 1914, Relf, using circular cylindrical obstacles, observed a transition for $Re$ bigger than 100-110 and Richardson observed one for $Re$ around 30, but unlike these authors, Bénard made this question a special object of research. It was not the case for Camichel's group in Toulouse, who performed precise experiments[51], with circular cylindrical bluff bodies obtaining the value of $Re_c$ = 47.

---

[30] Other times, he refers to the transition to from the "Poiseuille" regimen to the "hydraulic" regime.

This last determination persists today as the accepted value of the onset of the transition or bifurcation from the stationary flow state to the periodical instationnary flow state, in the case of a long body of circular cross section, far from the container's walls.

In terms of dynamical systems, the existence of an onset with a supercritical transition can be modelled by Andronov-Poincaré instability. Its normal form (or complex Landau equation[72]) describing the properties of the limit cycle of frequency $\omega$, appearing for positive values of a control parameter $\mu$, with the complex amplitude $A = \rho \exp(i\phi)$, where is $\rho$ the strength of the oscilation, can be expressed by the equation:

$$dA/dt = (\mu + i\,\omega_0)\,A - (g_1 + ig_2)\,A\,|A|^2 \qquad \text{with } g_1 > 0 \qquad (1)$$

or

$$d\rho/dt = \mu\rho - g_1\rho^3 \qquad (2a)$$

$$d\phi/dt = \omega_0 - g_2\rho^2 \qquad (2b)$$

This last equation is the law of frequency for a nonlinear oscillator, with a shift of frequency proportional to the square of the amplitude.

When a stationary process ($d\rho/dt = 0$) is established, from the real and imaginary components of the equation, the saturated amplitude is obtained, $\rho = (\mu/g_1)^{1/2}$ which varies with the square root of the growth rate $\mu$. If the control parameter of this instability (called also a Hopf bifurcation) is a Reynolds number, near the threshold at $Re_c$, $\mu$ is positive when $Re > Re_c$ et can expressed as linearly proportional to the distance of the critical value: $\mu \sim (Re - Re_c)$ [73].

In 1984, Christian Mathis, Michel Provensal and Louis Boyer [74] measured, by Doppler anemometry, the velocity fluctuations of the vortex shedding behind a cylinder and showed that the amplitude of these fluctuations were proportional to $(Re-Re_c)^{1/2}$. This allows them to verify the model given by equation *(1)* as well as the predictions of Landau-Stuart nonlinear models of instabilities in fluid dynamics [72]. Applying equation *(2b)*, the frequency grows linearly with the Reynolds number as

$$f = f_0 + B\,(Re - Re_c) \qquad (3)$$

with $f_0$ for the frequency at the onset, called the Hopf frequency. This is the frequency predicted by the complex eigenvalue by linear theory instability which is applied to the two-dimensional base flow $U(x,y)$, with $x$ ($y$) the streamwise (spanwise) coordinate, which is sensitive to the geometry of the section of the bluff body. Divided by the typical velocity $U$ and multiplied by the typical distance $l$, this law, in terms of the Strohual number, becomes

$$S = S_0 (1+ K/Re) \qquad (4)$$

exactly as was proposed by Rayleigh in 1915 in his famous law of similarity[31] [44]. As we see in the previous sections, the parameter $S_0 = (l^2/\nu) B$ is the asymptotically constant value of the Strouhal number when the Reynolds number goes to infinity. Historically, $S_0$ has been used for comparing experiments, as we reviewed before, and it depends on B, slope of the linear relation between the frequency and the Reynolds number. The coefficient $B \sim (g_2/g_1)$ measures the strength of the nonlinear (quadratic) correction of the frequency with the amplitude of the velocity fluctuations.

In order to carry out a retrospective analysis of the results of the pioneering experiments, we collected them in Figure 11, where the more significant of the old results we have been discussing are plotted all together, in the plan *St vs. Re*,. This diagram, introduced by Relf (Figure 10a) and used to verify similarity by many authors, shows sparse data, which explain why it is difficult to draw definite conclusions. But in reference to the Andronov-Poincaré instability model, we chose to verify directly the law of frequency with the Reynolds number (eq. 3). Plotting the non-dimensional frequency $F = Re. St = (l^2/\nu) f$, we clearly observe that almost all the results followed this linear law well, independently of the geometry of the body and of the fluid (Figures 12 and 13). For instance, we observe the Strouhal's results[45], analysed by Rayleigh [44] who postulated the law *F = 0,212 Re – 4,5,* as the fittest[32] for these  measurements.

We also analysed the measurements obtained before 1930, contemporaneous with Bénard, made on circular cylindrical bodies in air and in water by Relf [47],

---

[31] When the frequency is dimensionalised with the inverse of the viscous typical time *($l^2/\nu$)*, it is called the Roshko number $R_0$, [75-76] and the equation (*4*), becomes $R_0= R_{00} + A.Re$

[32] This law fits very well with more modern measurements, in the range *40<Re<150* [75]

Krüger[43], Richardson [49] in water and mixtures as well as Camichel [51]. We observe that they are not very far apart, following a linear law of frequency with the Re number. Exceptions are observed in the experimental data from Rubach and Kármán[42], with cylinders, and Richardson's results in air [49], which seem to be outside of the general trend .

As we said before, while in Lyon, Bénard produced a large quantity of precise frequency measurements of the vortex shedding in a wide range of velocity flows. From the fittest of 71 of his original results, we obtain the following law: *F= 0,25 Re – 24,41* (or *St= 0,25 (1-97,4/Re)*). Remember that Bénard observed [40] that, in contradiction with Kármán's, his values for the *St* number, change with the *Re* number[33].

These results verify a similarity law, as Rayleigh's one, but with different coefficients due to the fact that Bénard used prismatic cross section and no circular section. This is expected, because the value of the critical $Re_c$ number, the frequency at the onset and the nonlinear coefficients, are determinated by the two-dimensionnal velocity profile behind each different body [77].

Comparing these results, in the *S vs. Re* curves, with those obtained in 1921 by Relf [47] with cylinders, Bénard noted differences, leading in 1926, to the wrong statement: experiments did not follow a dynamical similitude law[34][78]

Two years later [55], Bénard compared his results with Camichel's, Dupin's and Teissié-Solier's [50] (Figure 10b) and corrected his statement, pointing out that the geometry of the bluff bodies and the streamlines around the bodies are not the same.

---

[33] In 1926[78], Bénard proposed the formula *f= ($l_1$U – v)/ $l_2$ (l+ $l_3$)*, with $l_1$, $l_2$ and $l_3$ typical lengths, as the synthesis of his different measurements, which also implies the law *S= $S_0$ (1-K/Re)*.

[34] The problem of the inexistence of an unique law *St= f(Re)* for all the alternated vortex in real fluids was discussed by Benard with Kármán, in the 2nd. Int. Congress of Appl. Mechanics in Zurich. Kármán attributed these variations to the existence of capillary problems originated in the free surface of the Benard's experiments. But the representation of the frequency in the space *S, Re* and the appropiate capillary parameter, the Weber number built with surface tension *We= (l $U^2$/σ)*, did not show a better similarity[79].

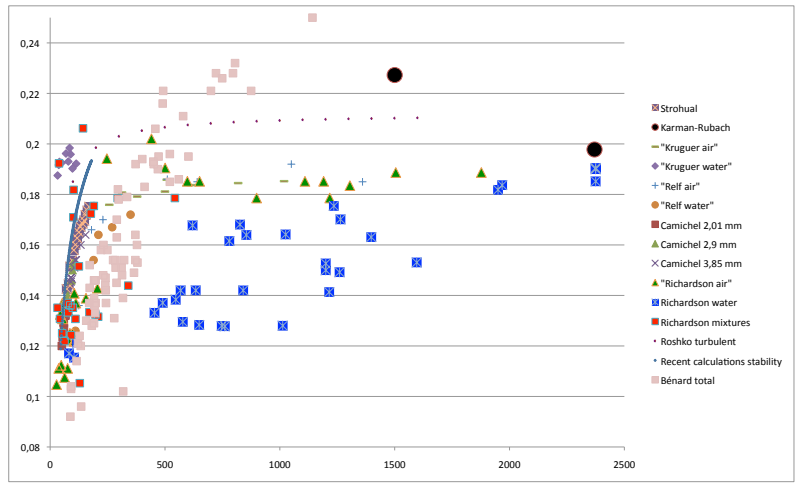

Figure 11: Plot of Strouhal number $S$ vs. the Reynolds number $Re$, measured in the different experiments discussed here.

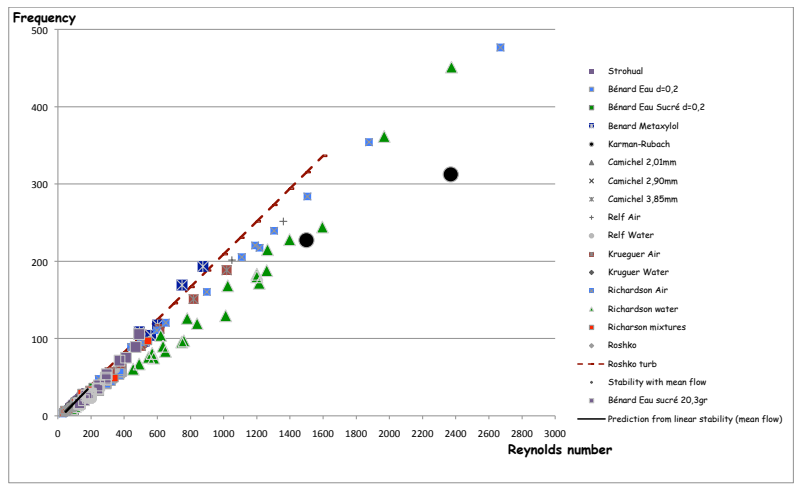

Figure 12: Plot of the value of the non-dimensional frequency $F = St.\ Re$ vs. the Reynolds number $Re$, measured in the different experiments discussed here.

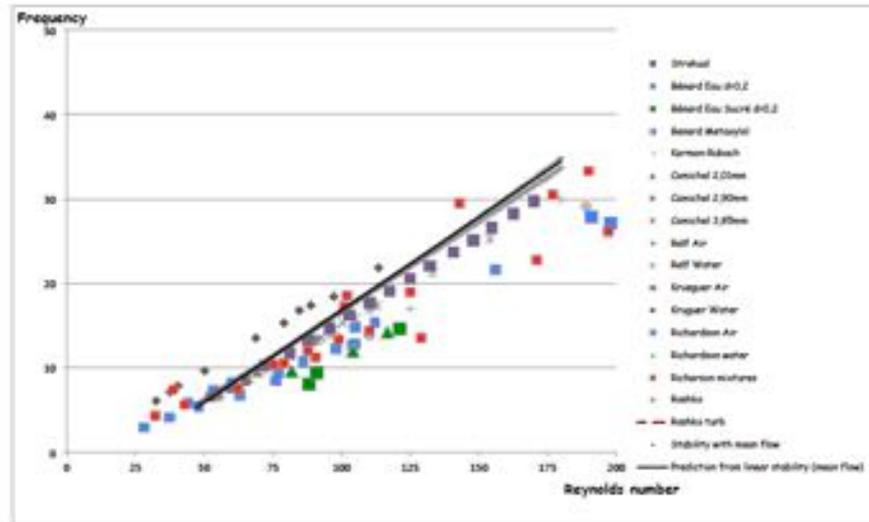

Figure 13: Same plot as in figure 12, but limited to Re < 180, where the vortex shedding behind circular cylinders is expected to be two-dimensional.

In conclusion, we recall that the law of frequency variation can be explained by the existence of a hydrodynamic instability, displaying an onset with finite frequency (Hopf instability). Because of the supercritical character of the instability, the amplitude of vortex velocity grows with the square root of the distance to onset, implying a linear variation of the frequency with the control parameter Re. Even if similarity laws were enounced by Rayleigh in 1915, mathematical explanation and experimental confirmation of the model were achieved only in 1984 by Mathis et al.[74].

In order to find predictions of the exact value of the observed frequency in the experiments, we have to wait until the early 2000s, when linear studies of instability were performed, but around the nonlinear modified mean flow profile ⟨$U(x,y)$⟩ around the circular cylinder. The existence of velocity fluctuations and quadratic nonlinearities of the Navier-Stokes equations, governing the description of the flows, generates a distortion of this average flow profile around the obstacle. As a consequence, its instability depends on this self-regeneration.

Recent calculations with this procedure, allows to obtain a new *S vs Re* curve [80-81-82-83](Figure 11), finally able to reproduce experimental results in a wide range of *Re* numbers.[35]

---

[35] These results are valid in the regime of two-dimensional instability (47<*Re*<180) occurring before the apparition of three-dimensional instabilities.

Taking these results into account, the best theoretical prediction for the dynamical similarity today, equivalent to the famous law of Rayleigh, becomes *S = 0,220 (1-23,01/Re)*

## 5. THE INSTITUTE OF FLUIDS MECHANICS IN PARIS AND CONCLUSIONS

This review of Bénard's scientific activity, in a historical context, illustrates his high capability to produce experimental work. A good summary of this aptitude was expressed in Jean Perrin's speech during the ceremony of the Socièté Française de Physique when Bénard was elected president: "He has given to hydrodynamics the methods of physics, and in doing so, he has discovered and studied phenomena not foreseen by theoreticians" as proved during the three stages of his scientific career.

During the first stage, at the Collège de France, as an autonomous young researcher preparing his PhD, he performed experiments that had a strong national and international impact. This work later will constitute a base for the development of theory of the hydrodynamic stability around the Rayleigh-Bénard convection. Even a century later, when theories about dissipative structures, spatio-temporal chaos and turbulence are being elaborated and many experimental facts developed around the subject, his original experiments are not outdated.

Few scientists before Bénard (for example Strohual or even Reynolds) were capable of accomplishing systematic and quantitative experiments that would eventually (more than half century later!) receive a theoretical explanation. It was the case with the thermal convection experiments of Bénard, which are now explained by the Marangoni effects produced by the variation of the surface tension with the temperature.

In addition, Bénard coupled his research activity on convection with original outreach. For instance in 1913, he produced with Dauzère, for the Gaumont company, several movies about the patterns generated in convection[84]. He also organized along with his collaborators, public exhibitions of his experiments in the Exposition de Physique et de T.S.F. at the Grand Palais in Paris in 1924 as well as for the World's Fair in 1937 in Paris.

The second stage of his career was dedicated to vortex shedding experiments and the analysis of very precise and numerous results obtained in Lyon by cinematographic means.

The discovery, from his experiments, of highly periodical vortex emissions, led to the definitive explanation of the Aeolian tone, consequence of the pressure variations induced by the vortex shedding behind wires or other obstacles. This subject motivated interest and debate around the verification of the existence of similarity laws or frequency laws. Comparing his work with Kármán's, Bénard insisted that his work was dealing with real flows, while Kármán's theoretical treatment was devoted to the stability of the alternate array of point vortex in ideal flows. At the same time Bénard was then fighting against the expression "von Kármán street" to designate the alternate vortex pattern, battling for the recognition of his early contribution to the description of the phenomenon .

It was necessary to wait nearly eighty years to understand the law of frequency (similarity laws) applied to hydrodynamic supercritical instabilities or Hopf bifurcation, like the vortex shedding behind bluff bodies. The exact value of the coefficients of the linear relationships between frequency and Reynolds numbers, near the onset, has been computed recently in case of vortex shedding behind circular cylinders. The value of the coefficients changes with the geometry of the cross section of the obstacle, even if the type of law is the same. This explains why Bénard, dealing with comparaisons of experiments with prismatic bodies and not with circular cylindes, changed his mind many times as to whether he was in agreement with the Rayleigh concept of similarity. At first, he wrongly believed that similarity implied the same law for all geometries, which explains why, when he moved to Paris, he decided to set up a new version of the Lyon experiments, but with a larger field of view[36]. This was so that he could observe more than five vortices in a 6 cm. field of view, as in the original experiments but with a higher rate camera, allowing higher precision in measuring the frequency. In 1928, Bénard, with the support of the Commission de la Journée Pasteur,[37] built a new experimental facility in Paris (Figure 14), equipped with a cinematographic camera Debrie G.V. at high rate (200 Hz) to obtain better temporal resolution for the measurements of the frequency.

---

[36] The new experiments, with using optics of a large diameter, specially built in the Observatoire de Paris by André Couder, allow for to the measurement of on 24 cm. of nearly of 20 vortices. As this, and, with a higher rate of views, the precision on frequency is increased[85].

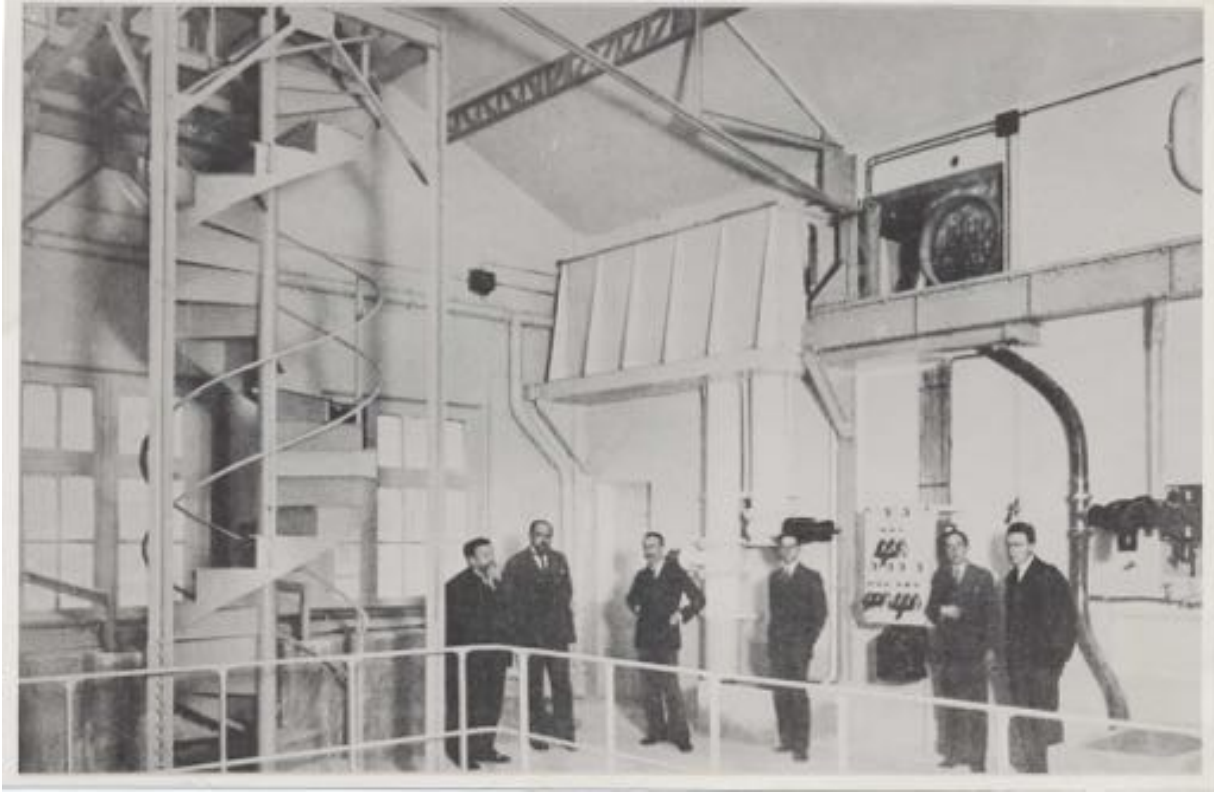

Figure 14 : Institut de Mécanique des Fluides de l'Université de Paris, on 4, rue de la Porte d'Issy- Paris XVe, in the 30's decade. From left to right: H. Bénard, D. Riabouchinsky, H. Villat, H. Journaud, C. (K) Woronetz (Voronetz) and L. Santon. On the left, the stairs to access to cinematographic camera used for the vortex shedding experiments.

The third stage of Bénard's scientific career was as Professor at the Sorbonne. Having acquired a great national reputation as a fine experimentalist he was elected in 1928 President of the French Physical Society and head of the experimental laboratory of the new Institut de Mécanique des Fluides at the Sorbonne, which had been heavily funded by the Air Ministry. He lived through the golden age of the institute during the 1930's, when new dependencies were built and when regular funding, as well as students, collaborators and visitors (Figure 15) arrived at the Institute. One the characteristics of this last period of Bénard's life was his open-minded approach to research orientations, with regards to geology, astronomy or atmospheric studies. He was always distant and reserved as a PhD thesis advisor, but he is remembered as "a delightful colleague", "always happy to help to young physicists who come to solicit his advice"[86]. Bénard almost always wrote the preface for his team's scientific publications[38] (but not the articles). However, he was also said to be modest to a fault, as he "disliked publishing and never presented a synthesis of his views" as one of his students, Paul Schwarz, said.

---

[37] Funds distributed by the Sciences Academy were gathered owing to a popular fundraising in order to support scientific activities for the centenary of Pasteur, organized in 1924, under Pasteur's motto *Sans laboratoires, les savants sont des soldats sans armes*.

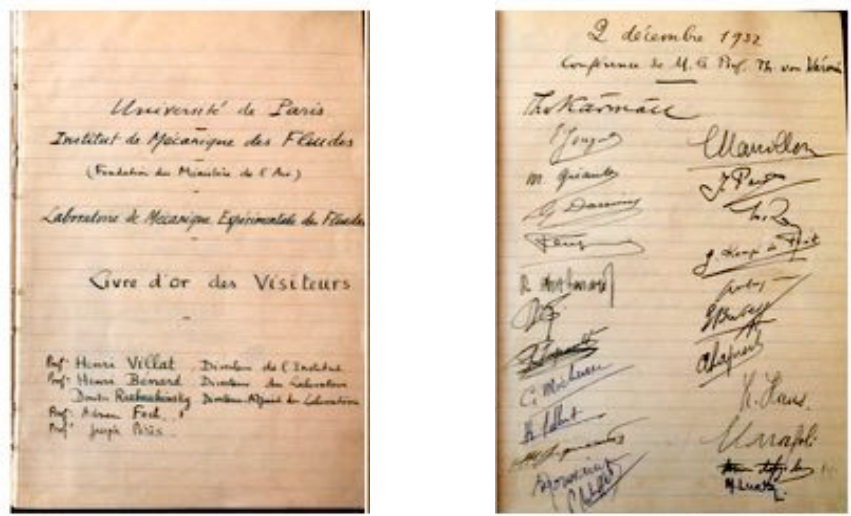

Figure 15: Golden book of visitors at the Institut de Mécanique des Fluides [ ]: a) cover page; b) participants' signatures of the Th. Kármán seminar on December 2nd. 1932, in the presence of Micheau, Giqueaux, Villat, Toussaint, Rebuffet, Barrillon, Pérez, Roy, Kampé de Feriet, Lapresle, Hans, Carafoli and Luntz amongst others[39].

---

[38] Usually, these publications are edited by the Air Ministery of the Air, in an important collection called the Publications Scientifiques et Techniques du Ministère de l'Air.

[39] It is odd to observe Bénard's absence. We have to remember the conflict between Bénard and Kármán around the naming of vortex streets. Kármán wrote the following [87]: *I never asked to have my vortex theory named after me, but somehow the name remained. There is always some danger in such matters, especially as one grows in fame or importance. In 1930, almost two decades after my paper was published, a French professor named Henri Benard popped up at an international congress and protested the name Kármán Vortex Street. He pointed out that he had observed the phenomenon earlier and that he had taken pictures of alternating vortices before I did. He was right, and as I did not wish to fight over names I said: "All right, I do not object if in London this is called Kármán Vortex Street. In Berlin, let us call it Karmansche Wirbelstrasse,and in Paris, Boulevard d'Henri Bénard. We all laughed heartily, and Bénard and I **became good friends*** Even if during these years Kármán entertained regular correspondence with other French scientists, especially with Charles Sandron, Philippe Wherlé and Joseph Kampé de Férat, but there is no evidence of letters between him and Bénard or any sign of friendship between them. The only document related to Bénard in Kármán's archives in Caltech (Pasadena), is a memorial card sent to Kármán when Bénard deceased in 1939.
After the war, Kármán became good friends with Henri Villat, Joseph Pérez, and Lucien Malavard and especially with Rolland Willaume who assisted Kármán in heading the technical NATO office in Paris, the AGARD [88], and in a joint consulting company, the BARA.

With Schwarz as a PhD. student, he came back to the vortex shedding experiments in the new facility in Paris, with the purpose of producing more precise values for the law of frequencies (Figure 14). This work finally focused on the study of the influence of the channel's transversal finite size and the asymmetry of the vortex street in non-symmetric channels. The effect of confinement on the vortex street stability was studied by Villat, who made it his main mathematical activity for a long period of time (see footnote 28). Villat was, at that time, the head of the Institute of Fluid Mechanics and he was probably responsible for the reorientation of Schwarz's work.

Bénard and another student, François-Joseph Bourrières[40], studied the fluid-structure interaction of an elastic pipe (a garden hose) subjected to running water. They followed its trajectory, thanks to a flashlight set at the free extremity of the hose, describing periodic motion and self-oscillations, for a set of flow conditions.

The phenomenon is considered similar to and "easier to study than the Bénard-Kármán vortex streets" [89]. Bourrières refers [90] to the "stability around dynamic trajectories" which is the limit cycle stability, evidencing knowledge about Aleksandr A. Andronov's studies[91] and other authors on the theory of dynamic systems in the late 1920s, and about the connection between Henri Poincaré's limit cycles and many systems showing oscillations [92].

Bénard wrote in the preface of Bourrières's work: "But we did not look for the non-periodic shape that the motion would have under arbitrary initial conditions . Nor did we look for the reason which, in case of self-oscillations, causes the system to tend more or less quickly towards a limit cycle which is identical whatever the initial conditions are".

From these comments, we think that Bénard was very close to interpret his experiments on the vortex shedding as a dynamical system showing a limit cycle, with onset[42].

---

[40] Bourrières (1880-1970) was a former undergraduate student of Bénard's and Duhem's in Bordeaux. In addition to the experimental work described, he established a partial differential equation as theoretical model for the fluid-structure interaction.

[42] This was an important concept for Andronov, who, following Poincaré, called a bifurcation value, a value of the parameter for which the phase portrait undergoes a qualitative change.

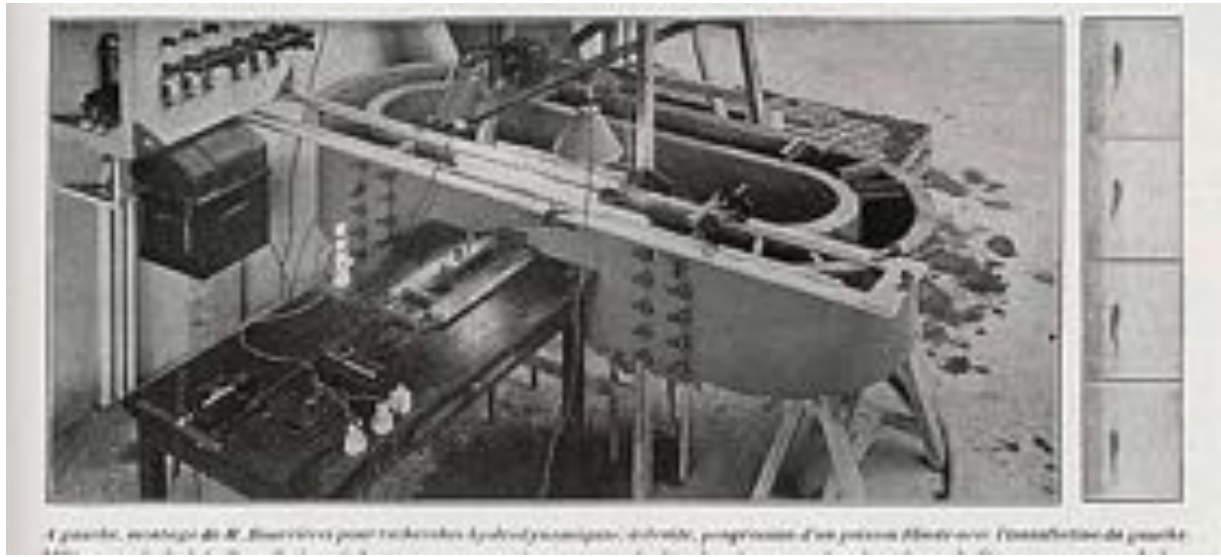

Figure 16 : Water channel (a) used by Bourrières, under Bénard's direction, to study undulating fish motions, by cinematography (b).

In an other experiment, Bourrières tried to compare the wavy periodical deformation of the elastic tube with other oscillations observed in nature like fish undulations (Figure 16), work that has never been published.

With his other students, Dusan Avsec and Michel Luntz, Bénard was interested in applications to meteorology and in the study of the convective structures in the form of rolls following the wind, in order to understand certain kinds of clouds. For this research, they collaborated with Philippe Wehrlé, Director of the Office National Météorologique (ONM) and Georges Dedebant, head of its scientific service [93]. In 1935, the Air Ministry created the Commission of the atmospheric turbulence, with Wherlé as Chairman. The commission was joined by P. Schereschewsky, J.-M. Kampé de Feriet, M. Kiveliovitch, J. Bass, J. Moyal, R. Failletaz, A. Giâo and J. Roulleau. Inspired by the theory of random variables and the statistical theory of turbulence, they developed a special hot-wire anemometer, attached to an airplane to register turbulent random fluctuations. This airplane was also used for the observation of cloud patterns in connection with the patterns observed in the laboratory by Bénard's group. The reports from the commission also mentioned "the discovery by Avsec and Luntz of the electro-convection, with application to the convective phenomena in the atmosphere, following Bénard's beautiful studies of the pre-turbulent states of transition between the laminar and the turbulent regime"[93].

In March 1939, Bénard passed away. Adrian Foch succeeded Bénard as professor,

assuming the continuity of his students' doctoral work.

The Institut de Mécanique des Fluides de Paris, which had previously merged with the Laboratory of Mechanics (solid mechanics) and the laboratory of Aeronautics of St Cyr, had been suffering from budget restrictions at the end of this decade. In 1940, the German army requisitioned the building and during the Occupation, the Air Ministry's scientific production of monographs was stopped. Wherlé describes the situation in September 1940: "Today, after the French defeat, I see all our enterprise seriously compromised, several of our collaborators and unfortunately among the best, placed under surveillance or forced to leave France because of their foreign nationality; our installations plundered and our material, even strictly scientific, became war looting and transferred to Germany"[94].

Many hydrodynamic scientists lost their position because of the discrimination against foreigners like Luntz[43], or because of racial laws against Jews like Jacques Valenci[44] in Marseille. The eradication of people and laboratories was ordered and imposed in the occupied zone, as in the case of Charles Sandron[45] in Strasbourg when the Faculty of Sciences moved to Clermont Ferrand. We will also mention Joseph Kampé de Fériet's case, Director of the Institut de Mécanique des Fluides in Lille. The Institute was transferred on 17th May 1940, along with 51 members, to Toulouse until

---

[43] Luntz (1909-?), born in Russia, was in the Institut de Mécanique des Fluides de Paris between 1931 and 1937 with Bénard. During 1938 and 1939, he worked at the Office National Métereologique, with Wherlé and Dedebant on turbulence, and subsequently was imprisoned by the French Vichy government in the camp of Vermont as a stateless person [95]. During his internment, he contributeds to mathematics theory on iteration, works communicated by Hadamard in [96], who denounced in his paper, Lutz's situation. Wherlé wrote to Kármán in Pasadena, asking for a position in the USA to help with his liberation. Luntz joined the ONERA after the Second World War

[44] Valenci, specialist in turbulence, lost his positions as Professor at the University of Aix-Marseille and at the Ecole de l'Air. In February 1941, he also contacted Kármán [94] asking for emigration to North or South America. Kármán wrote letters of support to the Rockefeller Foundation.

[45] Sandron was a pioneer in the study of hydrodynamics of polymeric solutions. Once in Clermont Ferrand, he joined the Resistance and was arrested by the Gestapo in November 1943 and deported to the Dora-Mittelbau camp of forced labour, dedicated to the production of spare parts for the V2 bomb. On one occasion, he came across Werner von Braun who was then visiting the camp. He made a proposition to Sandron to join his laboratory at Peenemunde. Sandron refused.[97]
Few years before, Sandron had written a book with Kármán on fluid mechanics. Villat offered his help for the publication in Gauthier-Villars. But the financial conditions were so high that Sandron had to renounce. Later on, he discovered that Pérès was about to publish a similar book. Sandron came to the conclusion that Villat had incited the publisher to impose to him an inacceptable price in order to prevent the publication of his book and favored the one of Pérès. Sandron, in a letter to Kármán dated 7 June 1936, denounced that Villat had "torpedoed" the book, which will never be published. He wrote: *Evidemment, tout ça est un peu dégoutant, mais vous ne devez pas oublier, mon cher Patron, que les hommes qui sont en ce moment influents dans l'Université française ne sont pas tous des caractères admirables, et que leurs petites combinaisons les intéressent beaucoup plus que la Science.*[94]

Liberation day[46]. One of these members was François N. Frenkiel[47], who was arrested by the Gestapo and remained in concentration camps until 1945. He emigrated to the United States after the war and in 1958, became the first Editor of Physics of Fluids. After the war, one part of the French fluid mechanics was reorganised, reinforced with the CNRS, with the creation of the CEA (atomic energy institution) and specifically with the creation of the ONERA (aeronautical research centre) with strong relationships with the aeronautic and the defence industries. The former members of the Institut de Mécanique des Fluides of Paris moved to new centres like Orsay, in the new center built by the Faculty of Sciences or to the Ecole Normale Supérieure. In the CNRS, part of the discipline was concentrated in the section "Mécanique générale et mathématiques appliquées" (section 3) created in 1948 by Joseph Pérès, at that time deputy director of the CNRS (1945-49).

The subjects developed by Bénard were less followed. But a renewal of interest appeared first in the 1960's, around studies on hydrodynamic stability related to astrophysics and plasma physics, with S. Chandrasekhar's book [98], who gave a full overview of the Rayleigh-Bénard convection[48]. At that time, Paul Glansdorff's and Ilya Prigogine's work [100] about thermodynamics out of equilibrium also revived the role of Bénard in the discovery of the well organised "dissipative structures" obtained under non equilibrium external heat flow. But in the 1970's, physicists of condensed matter and phase transitions, interested in the universality concepts of the critical phenomena, explored new fields of application of these ideas and drew attention to transitions between laminar and unstable states in fluid dynamics. With this impulsion, an intense activity was developed around subjects like the Rayleigh-Bénard convection, the Bénard- Marangoni convection, the Taylor-Couette centrifugal instability, the Frederiks transition in liquid crystals, the Belousov Zhabotinsky chemical reaction, the Bénard-von Kármán vortex shedding instability and, later, the subcritical transition to turbulence in confined shear flows. At the same time, theories of bifurcations, and catastrophes, dynamical systems and nonlinearities are redynamised in mathematics.

---

[46] During the first two years in Toulouse, they worked quietly (living in an old castle, in the middle of a magnificent park) and were able to research turbulence and atmospheric boundary layers. But, after November 1942 (German occupation of the south of France), the situation became very dangerous to carry on with fluid mechanics research without the Germans' agreement. So the laboratory was hidden in an old church and the Germans never found it; the wind tunnel was in the choir and the hot-wire apparatus on the organ loft. Kampé wrote: *Toulouse was full of Germans, but none had the idea to look what was done in this old building! As the German troops were driven out of Toulouse (19 August 1944) they set fire to our block; all houses in the block were burnt to the ground; but our old church stood, proud, in the fire and, wonder, our laboratory is still now untouched and unharmed in the ruins*. [94]

[47] Kampé described this situation[94, M. Eckert private communcation]: *M. Frenkiel is of Polish nationality and during the German occuoation he has flown with myself to Toulouse...Mr. Frenkiel was not able to get the French exit visa. ... when the German troops came to Toulouse, I succeeded two times to avoid for him the German camp of concentration, but in April 1943 Mr. Frenkiel, believing to be better in the Italien zone, went to the Alps, after the Italian surrender he was caught in Italy with Mrs. Frenkiel by the German Gestapo and send to Auschwitz and Buchenwald. He was liberated after terrible experience by the Vth Army in April 1945*. (JEW: His pregnant first wife was died in one of the camps).

[48] In an interview with Spencer Weart [99], Chandrasekhar explained his interest on Bénard convection, He worked closely with experimenters in the Chicago University's campus, as Dave Fultz and Yoshinari Nakagawa working on hydrodynamics instabilities. With Enrico Fermi, as a consequence of his interest in Bénard convection, they studied gravitational stability in presence of magnetic fields.

In France, Pierre Gilles de Gennes's lectures at the Collège de France, especially in 1974 and 1975, were devoted to the subjects of hydrodynamics and generated strong interest among experimentalists and theorists, as many of them devoted themselves to these old subjects, but with new impulse. Previously, in 1973, a session of the Ecole de Physique de Les Houches[101-102] gathered people coming from the fluid mechanics and physics communities, who were playing an important role in the formation of a new generation of scientists. They specifically returned, but with modern experiments, to Bénard's subjects nearly a century later, as we described in this paper. Even if the more recent history of this subject still needs to be written[49], it could be said that during this more recent period, the interaction between theory and experiment was more sustained and regular. The misunderstanding suffered by Bénard is no longer present.

ACKNOWLEDGEMENTS

We would like to thank for their assistance Dwigth Barkley, Olivier Darrigol, Jan Dusek, Michael Eckert, Marianne Kosling, Laurent Limat, Anthony Pearson and Yves Pomeau.

---

[49] Recent works about the history of the development of the nonlinear science and chaos in France, can be red in [92] and [103]